\documentclass[fleqn,usenatbib,useAMS]{mnras}

\usepackage{graphicx}	% Including figure files
\usepackage{amsmath}	% Advanced maths commands
\usepackage{amssymb}	% Extra maths symbols
\usepackage{multicol}        % Multi-column entries in tables
\usepackage{bm}		% Bold maths symbols, including upright Greek
\usepackage{pdflscape}	% Landscape pages

\newcommand{\cmlt}{$\alpha_{\rm MLT}$} 
\newcommand{\ccm}{$\alpha_{\rm CM}$} 
\newcommand{\sad}{$s_{\rm ad}$}
\defcitealias{Spada_ea:2018}{Paper I}
\defcitealias{Spada_Demarque:2019}{Paper II}

\def\araa{{ARA\&A}}             % Annual Review of Astron and Astrophys
\def\apj{{ApJ}}                 % Astrophysical Journal
\def\apjl{{ApJ}}                % Astrophysical Journal, Letters
\def\apjs{{ApJS}}               % Astrophysical Journal, Supplement
\def\aap{{A\&A}}                % Astronomy and Astrophysics
\def\apss{{Ap\&SS}}          % Astrophysics and Space Science
\def\mnras{{MNRAS}}             % Monthly Notices of the RAS
\def\ssr{{Space~Sci.~Rev.}}
\usepackage[T1]{fontenc}
\usepackage{ae,aecompl}

\usepackage{newtxtext,newtxmath}
\title[Entropy-calibrated mixing-length for red giants]{
Stellar evolution models with entropy-calibrated mixing-length parameter: application to red giants
}
\author[Spada, Demarque, and Kupka]{Federico Spada$^{1}$
\thanks{Contact e-mail: \href{mailto:spada@mps.mpg.de}{spada@mps.mpg.de}},
Pierre Demarque$^{2}$
and Friedrich Kupka$^{1,3,4}$ \\
$^{1}$Max-Planck Institut f\"ur Sonnensystemforschung, Justus-von-Liebig Weg 3, 37077 G\"ottingen, Germany\\
$^{2}$Department of Astronomy, Yale University, New Haven, CT 06520-8101, USA \\
$^3$Univ. of Appl. Sciences Technikum Wien, Dept. Applied Mathematics and Physics, H\"ochst\"adtplatz 6, A-1200 Wien, Austria \\
$^4$Wolfgang-Pauli-Institute c/o Faculty of Mathematics, University of Vienna, Oskar-Morgenstern-Platz 1, A-1090 Wien, Austria
}

\date{Last updated ; in original form}
\pubyear{}

\begin{document}
\label{firstpage}
\pagerange{\pageref{firstpage}--\pageref{lastpage}}
\maketitle
%
%%%%%%%%%%%%%%%%%%%%%%%%%%%%%%%%%%%%%%%%%%%
\begin{abstract}
% Note: must be < 250 words long
We present evolutionary models for solar-like stars with an improved treatment of convection that results in a more accurate estimate of the radius and effective temperature.
This is achieved by improving the calibration of the mixing-length parameter, which sets the length scale in the 1D convection model implemented in the stellar evolution code.
Our calibration relies on the results of 2D and 3D radiation hydrodynamics simulations of convection to specify the value of the adiabatic specific entropy at the bottom of the convective envelope in stars as a function of their effective temperature, surface gravity and metallicity.
For the first time, this calibration is fully integrated within the flow of a stellar evolution code, with the mixing-length parameter being continuously updated at run-time.
This approach replaces the more common, but questionable, procedure of calibrating the length scale parameter on the Sun, and then applying the solar-calibrated value in modeling other stars, regardless of their mass, composition and evolutionary status.
The internal consistency of our current implementation makes it suitable for application to evolved stars, in particular to red giants.
We show that the entropy calibrated models yield a revised position of the red giant branch that is in better agreement with observational constraints than that of standard models.
\end{abstract}

%%%%%%%%%%%%%%%%%%%%%%%%%%%%%%%%%%%%%%%%%%%

%%%%%%%%%%%%%%%%%%%%%%%%%%%%%%%%%%%%%%%%%%%%
\section{Introduction}
\label{introduction}
%%%%%%%%%%%%%%%%%%%%%%%%%%%%%%%%%%%%%%%%%%%%

In stars with effective temperature $T_{\rm eff} \lesssim 10^4$ K, an outer convection zone develops because of the large local value of the opacity (e.g., \citealt{KWW12}).
The details of the formalism adopted to describe convection are of critical importance in modeling these stars.
In particular, special care must be taken in modelling the outer convective layers, close to the stellar photosphere.

When convection occurs in a star in sufficiently deep layers, it is very effective at transporting energy, and is therefore amenable to a simple description, based on the assumption of an essentially adiabatic stratification.
Closer to the stellar surface, in contrast, convection becomes less and less efficient, due to the decreasing local density, and the increasing radiative loss from the convective elements.
As a result, a significantly super-adiabatic stratification typically develops in the subsurface layers of a stellar convection zone \citep[e.g.,][]{Kupka:2020}.
The details of this super-adiabatic layer (SAL) determine the specific entropy in the adiabatic part of the convection zone, $s_{\rm ad}$, which in turn controls the radius of the star \citep[see, e.g.,][for details]{Schwarzschild:1958, HansenKawalerTrimble:2004}.

No fully satisfactory description of the SAL that can be incorporated in 1D stellar evolution codes exists to date. 
Stellar models thus routinely rely on first-order, local models of turbulent convection.
Perhaps the most commonly adopted of such models is the mixing-length theory \citep[MLT;][]{BV58,Cox_Giuli:1968},  of which several equivalent, but slightly different formulations exist (see Appendix A of \citealt{Ludwig_ea:1999} for a concise summary). 
Some of the assumptions of the MLT are, however, not satisfied in real stars \citep[see, e.g.,][for details]{Trampedach:2010}.
In an effort to overcome some of the shortcomings of MLT, an improved model was developed, which is also widely used in stellar evolution codes \citep[CM in the following]{Canuto_Mazzitelli:1991,Canuto_Mazzitelli:1992,Canuto_ea:1996}.

A specific issue shared by the  MLT and the CM formalisms is that they require the use of a free parameter to set the length scales of turbulent eddies in the model (\cmlt{} and \ccm{}, in the following, respectively), thus requiring an external calibration.
In the most common approach, the Sun is adopted as a calibrator for \cmlt{} (by constructing a standard solar model: \citealt{Gough_Weiss:1976}; see, also \citealt{Basu_Antia:2008}), and then this solar-calibrated value is assumed to be also applicable to modelling other stars.
This assumption, however, is not theoretically justified, and there is increasing observational evidence that \cmlt{} indeed varies from star to star (e.g., \citealt{Lebreton_ea:2001, Yildiz_ea:2006, Bonaca_ea:2012, Joyce_Chaboyer:2018, Viani_ea:2018}, and references therein).

An alternative approach to the solar calibration of the \cmlt{} parameter is clearly desirable, on both theoretical and observational grounds.
A promising avenue towards this goal is to exploit the results of numerical simulations of convection, implementing a realistic treatment of radiation hydrodynamics (RHD)\footnote{The simulations we consider in this work are also formally referred to as ``large eddy simulations'' (as opposed to ``direct numerical simulations''), since they introduce some viscosity model to account for the influence of unresolved scales on the flow \citep[see][for a review]{Kupka_Muthsam:2017}.}.
Indeed, since the first RHD simulations of stellar convective envelopes became available, there have been attempts to incorporate their results in stellar evolution codes \citep[e.g.,][]{Lydon_ea:1992,Lydon_ea:1993a,Lydon_ea:1993b}.

Within the framework of an MLT-like description of convection, a systematic approach to the calibration of the length scale parameter was developed in the seminal works of \citet{Steffen:1993}, \citet{Ludwig_ea:1997}, and \citet{Freytag_ea:1999}, based on the picture of convection already emerging from the early 2D and 3D RHD simulations \citep[e.g.,][]{Stein_Nordlund:1989}.
The key idea of this method is that although the details of the stratification profile obtained from RHD simulations cannot be reproduced by the MLT formalism, it is possible to select a value of \cmlt{} that results in a 1D stellar envelope having the same adiabatic specific entropy of the simulations. 
In this sense, the MLT parameter can be calibrated across the HR diagram if the value of \sad{} as a function of effective temperature, surface gravity, and metallicity is available from the RHD simulations.

Based on the entropy-matching approach just outlined, several authors have constructed calibrations of the mixing parameter of convection using RHD simulations.
 \citet{Ludwig_ea:1999} produced a calibration of both \cmlt{} and \ccm{} based on their 2D RHD simulations, constructed with the code described in \citet{Ludwig_ea:1994}.
Using the calibration of \cmlt{} obtained by \citet{Ludwig_ea:1999}, \citet{Freytag_Salaris:1999} constructed evolutionary models of globular cluster stars.
More recently, \citet{Magic_ea:2015a} used the same technique of \citet{Ludwig_ea:1999} to calibrate \cmlt{} using the 3D simulations from the \texttt{STAGGER} grid \citep{Magic_ea:2013a}.
\citet{Trampedach_ea:2014b} performed an analogous calibration of \cmlt{} with an independent method, based on the 3D RHD simulations of \citet{Trampedach_ea:2013}.
The latter calibration was used for the first time to construct stellar evolution models by \citet{Salaris_Cassisi:2015}.
Finally, \citet{Sonoi_ea:2019} updated the original calibration of \cmlt{} and \ccm{} of  \citet{Ludwig_ea:1999} using 3D RHD simulations from the CIFIST grid \citep{Ludwig_ea:2009}, constructed with the \texttt{CO$^5$BOLD} code \citep{Freytag_ea:2002}.

Our approach to the calibration of \cmlt{} shares the same theoretical foundation of that originated with \citet{Steffen:1993}, but, following \citet{Tanner_ea:2016}, differs in one key methodological aspect.
We integrate the calibration of \cmlt{} directly within the stellar evolution code, taking as input the values of \sad{} from the RHD simulations.
The MLT parameter is updated continuously along the evolutionary track, thus acquiring a dependence on the surface conditions of the star (effective temperature, surface gravity, and metallicity).
In our approach, the adiabatic specific entropy replaces the mixing length parameter as the main calibration variable.
From the stellar modelling point of view, this shift has significant theoretical and practical advantages.
The mixing length parameter is primarily a modelling variable, which can only be loosely interpreted as a measure of the efficiency of convection, and is notoriously sensitive to the details of the input physics, and even to the numerics of the stellar evolution code.
In contrast, \sad{} is a physical quantity, whose dependence on stellar parameters is much more robust and predictable.

This is the third paper in a series focusing on the application of the entropy calibration method of \cmlt{} suggested by \citet{Tanner_ea:2016} to construct stellar evolution models.
\citealt{Spada_ea:2018}, hereafter \citetalias{Spada_ea:2018}, presented the first application of the method to the construction of solar models.
The second paper (\citealt{Spada_Demarque:2019}, \citetalias{Spada_Demarque:2019} in the following) discussed the application to $\alpha$ Centauri A and B; showing that the entropy calibration improves the accuracy of the radii and effective temperatures with respect to the solar calibration.

Both previous papers used a preliminary implementation of the method, in which an external driver script handled the entropy calibration of \cmlt{}, and called the stellar evolution code to execute the next evolutionary step.
We found that this approach is adequate to model stellar evolution from the early pre-main sequence up to the sub-giant phase, but it is unsuitable for red giant branch evolution.
In this paper, we describe an upgraded implementation, in which for the first time the entropy calibration is an integral part of the stellar evolution code.
The enhanced internal consistency makes the code suitable to model evolved stars, in particular red giants.

This paper is organised as follows.
The details of the entropy calibration of \cmlt{} within the stellar evolution code are presented in Section~\ref{methods}. We discuss the impact of the calibration on the evolutionary tracks and on the internal profile of the entropy in Section~\ref{results}; a comparison with the observations is also shown.
Our results and conclusions are discussed in Section~\ref{discussion}, and summarised in Section~\ref{conclusions}.

%%%%%%%%%%%%%%%%%%%%%%%%%%%%%%%%%%%%%%%%%%%%
\section{Methods}
\label{methods}
%%%%%%%%%%%%%%%%%%%%%%%%%%%%%%%%%%%%%%%%%%%%

In the following, we present stellar evolution models calculated implementing a calibration of the mixing length parameter based on the results of RHD simulations of convection.
The approach was conceptually developed by \citet{Tanner_ea:2016}. 
In the previous papers of this series, \citetalias{Spada_ea:2018} and \citetalias{Spada_Demarque:2019}, we used this method to construct models for the Sun and for $\alpha$ Centauri A and B, respectively.
In this paper we present models constructed with an entropy calibration of \cmlt{} fully integrated within the flow of the stellar evolution code.

\subsection{The stellar evolution code}
\label{inputphysics}

All the models were constructed with an opportunely modified version of the Yale stellar evolution code, YREC \citep{Demarque_ea:2008}.
Unless otherwise specified, the details of the input physics implemented in the models are as follows.
We use the OPAL 2005 equation of state \citep{Rogers_Nayfonov:2002}, and the OPAL opacities \citep{Rogers_Iglesias:1995,Iglesias_Rogers:1996}, supplemented by the low-temperature opacities of \citet{Ferguson_ea:2005} at $\log\, T \leq 4.5$. 
Diffusion of helium and heavy elements is taken into account according to the formulation of \citet{Thoul_ea:1994}. 
In the atmosphere, the Eddington grey $T$--$\tau$ relation is used. 
Abundances of elements are scaled to the \citet{Grevesse_Sauval:1998} solar mixture, with $(Z/X)_\odot=0.0230$. 
The treatment of convection in YREC follows the MLT formalism (\citealt{Henyey_ea:1965}; see also 
\citealt{Paczynski:1969}). 
Our evolutionary tracks begin from starting models on the birth line \citep[see][for details]{Spada_ea:2013}.
For reference, with the above choices of input physics the standard solar model calibration of \cmlt{} in our code yields a value of $\alpha_\odot=1.8376$, and proto-solar (birth line) chemical composition $X_0 = 0.70615$, $Z_0 = 0.01885$.

In the following we also discuss standard tracks constructed using the atmospheric $T$--$\tau$ relation of \citet{KrishnaSwamy:1966}, all the remainder input physics being the same as above. 
In this case, the solar calibration yields: $\alpha_{\rm MLT}=2.1537$, $X_0 = 0.70618$, $Z_0=0.01884$.

Optionally, as an alternative to the MLT, the description of convection according to the \citet{Canuto_Mazzitelli:1991} ``full spectrum turbulence'' model (CM in the following) is available, following the implementation of \citet{Stothers_Chin:1995}.
{In this formulation, the scale length $\Lambda$ is set, similarly to the MLT approach, by $\Lambda = \alpha_{\rm CM}\, H_P$, where $H_P$ is the local pressure scale height.
}

\subsection{The adiabatic specific entropy from RHD convection simulations}

Our implementation of the entropy calibration of \cmlt{} requires that the value of the adiabatic specific entropy, determined from a set of RHD numerical simulations of convection, is specified as a function of the effective temperature,  $T_{\rm eff}$, surface gravity, $\log\,g$, and surface composition of the model.
The calibration procedure, performed at the beginning of each evolutionary step, evaluates the difference between this target value of $s_{\rm ad}$ and the current specific entropy of the model, and adjusts \cmlt{} to drive the difference to zero (see Section \ref{implementation} for details).

In practice, the RHD simulations are used to map the function:
\begin{equation}
\label{rhdentr}
s_{\rm ad}^{\rm RHD} = f(T_{\rm eff}, \log \, g; {\rm [Fe/H]}). 
\end{equation}
Both \citet{Ludwig_ea:1999} and \citet{Magic_ea:2013a} conveniently provided this functional relation in analytical form, derived from a best-fit of their respective simulations (see Appendix \ref{rhdfits}). 
The best-fits are functions of the two variables $T_{\rm eff}$ and $\log\,g$; in \citet{Magic_ea:2013a}, several sets of best-fitting coefficients are given, one for each of the metallicities available in their grid.
In both cases, the functional forms of the best-fits were chosen on purely numerical grounds, and do not contain any special physical meaning.
In order to assess the impact of the choice of the \sad{} function in equation \eqref{rhdentr} on the entropy-calibrated models, we have used both the \citet{Ludwig_ea:1999} and the \citet{Magic_ea:2013a} fits.

\begin{figure}
\begin{center}
\includegraphics[width=0.49\textwidth]{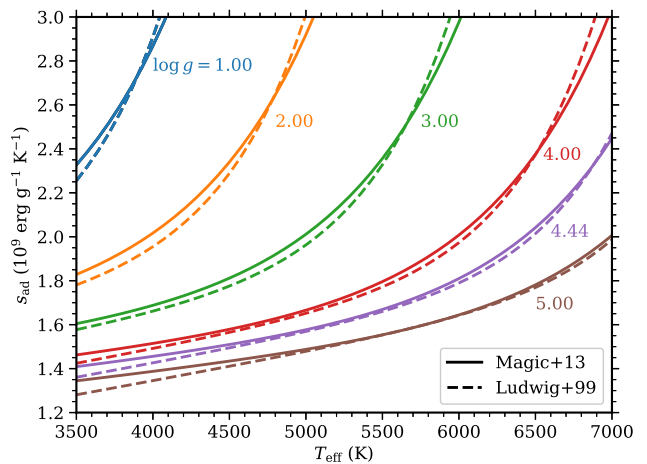}
\caption{Adiabatic specific entropy as a function of effective temperature and surface gravity from RHD simulations used in our entropy calibration. Solid lines: 3D simulations of \citet{Magic_ea:2013a}; dashed: 2D simulations of \citet{Ludwig_ea:1999}.}
\label{sbot}
\end{center}
\end{figure}

The \sad{} fits used in this paper are plotted in Figure \ref{sbot}.
We note that the functions are in good qualitative and quantitative agreement in the range of effective temperature and surface gravity typical of the main sequence, while more significant discrepancies arise for $T_{\rm eff} \lesssim 4500$ K and/or $\log\, g \lesssim 3.0$.
We can therefore expect the largest differences in the entropy-calibrated models to occur in the low $T_{\rm eff}$, low $\log\, g$ regime.

Our present choice of \sad{} differs from the one implemented in the models of \citetalias{Spada_ea:2018} and \citetalias{Spada_Demarque:2019}, where the functional relation in equation \eqref{rhdentr} was specified according to the best-fits developed by \citet{Tanner_ea:2016}, based on the simulations of \citet{Tanner_ea:2013a}, \citet{Tanner_ea:2013b}, \citet{Tanner_ea:2014}, and the grid of \citet{Magic_ea:2013a}.
We have verified that the fitting function of \citet{Tanner_ea:2016} and of \citet{Magic_ea:2013a} are essentially equivalent as far as our entropy calibration models are concerned.

\subsection{Implementation within the stellar evolution code}
\label{implementation}

\begin{figure}
\begin{center}
\includegraphics[width=0.49\textwidth]{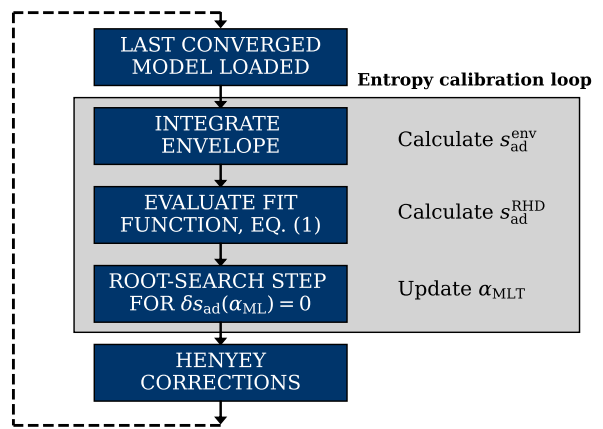}
\caption{Schematic flow of the entropy calibration procedure integrated in YREC. The calibration routines are only called at the beginning of a new evolutionary step and do not interfere with the rest of the program, except for providing the value of \cmlt{}.}
\label{flowchart}
\end{center}
\end{figure}

The entropy calibration of \cmlt{} is performed at the beginning of each evolutionary time step.
Figure \ref{flowchart} outlines the integration of the entropy calibration in the YREC code.

The calibration takes the form of an iterative procedure analogous to that described in \citetalias{Spada_ea:2018}.
Each iteration consists of the following steps:
\begin{enumerate}
\item A stellar envelope integration is performed with the current value of the mixing length parameter, obtaining the corresponding value of the adiabatic specific entropy in the model, $s_{\rm ad}^{\rm env}$;
\item The function in equation \eqref{rhdentr} is evaluated for the current $T_{\rm eff}$ and $\log\,g$, and the difference $\delta s_{\rm ad} = s_{\rm ad}^{\rm env} - s_{\rm ad}^{\rm RHD}$ is calculated;
\item The trial value of \cmlt{} is updated using a root-finding method to drive $\delta s_{\rm ad}$ to zero, and the loop starts over.
\end{enumerate}
The exit condition from the loop requires that either the absolute difference $|\delta s_{\rm ad}|$ or the absolute change in \cmlt{} with respect to the previous iteration fall below a pre-defined tolerance.
We used a tolerance on $\delta s_{\rm ad}$ of $5\cdot 10^{-4}$, and of $10^{-6}$ on \cmlt{}.  

{
Our root-search procedure for the equation $\delta s_{\rm ad}(\alpha_{\rm MLT}) = 0$ implements the ``Pegasus" root-finding algorithm, an improved version of the classical secant method described by \citet{Dowell_Jarratt:1972}.
Convergence to the root within the tolerances specified above is achieved in $\approx 3$--$5$ iterations, depending on the evolutionary phase.
Since these iterations consist of an envelope integration only, the performance improvement compared to the version of the entropy calibration used in \citetalias{Spada_ea:2018} and \citetalias{Spada_Demarque:2019} is considerable.
Two initial guesses for \cmlt{} are required to start the method.
These are reset at the beginning of each evolutionary step to ensure that the root is bracketed between them. }
Once the \cmlt{} calibration is performed, the program continues in its normal flow, described elsewhere \citep{Demarque_ea:2008}.

Beside the significant performance improvement, the integration of the entropy calibration within the flow of the stellar evolution code has several advantages over the approach used in \citetalias{Spada_ea:2018} and \citetalias{Spada_Demarque:2019}. 
The calibration procedure is a part of the envelope integration, and does not interfere with the rest of the code. 
In addition, the calibration routines have access to the parameters of the current stellar model, thus ensuring full internal consistency.
For the application to the post-main sequence evolution these improvements are not trivial, since we found that a correction factor depending on the mean molecular weight of the envelope needs to be applied to equation \eqref{rhdentr}, as discussed in the next Section.

\subsection{Correction factor arising from the change in the mean molecular weight}
\label{factor}

To construct entropy-calibrated models of red giant stars accurately, we found it necessary to introduce a correction factor to the value of \sad{} given by equation \eqref{rhdentr}. 
This factor takes into account the difference in the mean molecular weight between the model and the RHD simulations.
The physical meaning of this correction is the following.

The chemical composition used in the RHD simulations is determined by the choice of metallicity and helium fraction, independently of the location of the modelled star in the HR diagram.
The simulations of \citet{Ludwig_ea:1999} were performed at solar metallicity (with pre-\citealt{Asplund_ea:2005} solar mixture) and with initial helium fraction $Y=0.28$, while the simulations of \citet{Magic_ea:2013a,Magic_ea:2015a} were performed at different metallicities, assuming the \citet{Asplund_ea:2009} abundances, and a fixed value of the helium fraction (see the respective papers for details).

In a stellar evolution model, on the other hand, the surface chemical composition changes with time. 
Composition changes are driven by the nuclear reactions occurring in the core, and propagate to the surface through mixing by convection and gravitational settling.
As far as the calculation of the entropy of the model is concerned, these changes are quite small and safely negligible during the pre-main sequence, the main sequence, and the sub-giant phases, but they become quite large at the beginning of the red giant phase, when the convection zone deepens significantly and the so-called ``first dredge-up'' occurs. 
At this time, deep layers enriched in helium by the nuclear reactions in the core are partially mixed with the surface layers, which still have almost pristine composition (apart for the effect of gravitational settling).
As a consequence, the helium content of the outer layers changes by $\approx 25\%$ within a relatively short time scale ($\lesssim 1$ Gyr for $M \lesssim 1.2 \, M_\odot$).
This results in a discrepancy in the mean molecular weight between the model and the RHD simulations used to calibrate \cmlt{} that cannot be neglected.

To account for the evolution of the mean molecular weight, we exploited the fact that the specific entropy of an ideal gas can be written as \citep[e.g.,][]{HansenKawalerTrimble:2004}:
\begin{equation*}
s = \frac{\cal R}{\mu} \ln \left(\frac{T^{3/2}}{\rho}\right),
\end{equation*}
where $\cal R$ is the universal gas constant, $\mu$ is the mean molecular weight of the gas, and $T$ and $\rho$ are its temperature and density, respectively.
Even if this expression is not exactly valid for the gas in the stellar envelope due, for instance, to the effect of partial ionisation, we can assume that the scaling $s \propto \frac{1}{\mu}$ holds.
Based on this assumption, we introduced the following correction to the adiabatic specific entropy from the RHD simulation:
\begin{equation}
\label{mufactor}
f_\mu = \frac{\mu({\rm RHD})}{\mu({\rm model})},
\end{equation}
where $\mu({\rm RHD})$ and $\mu({\rm model})$ are the mean molecular weights calculated at the chemical composition of the RHD simulations and of the current stellar model, respectively.
The main contribution to the deviation of $f_\mu$ from unity is due to the difference between the helium fraction in the model and in the RHD simulations.
A smaller effect, also taken into account by $f_\mu$, arises from the difference in the metal fraction, due to the different assumptions on the solar mixture, which determines $(Z/X)_\odot$, and hence $Z$.
To ensure internal consistency, we calculate $\mu({\rm RHD})$ and $\mu({\rm model})$ through a call to the routine implementing the OPAL 2005 EOS used in the rest of the code.
While $\mu({\rm RHD})$ is fixed for a given evolutionary track once the set of RHD simulations has been selected, $\mu({\rm model})$ is updated at each time step with the current values of the composition parameters in the envelope. 
In the entropy calibration procedure outlined above, the value of $s_{\rm ad}^{\rm RHD}$ is given by equation \eqref{rhdentr} multiplied by the factor $f_\mu$ from equation \eqref{mufactor}.

It should be noted that the effect of the gravitational settling on the mean molecular weight of the envelope is comparable in size to that of the dredge up, but occurring over a longer time scale, essentially equal to the duration of the main sequence, i.e., several Gyr.
Of course, if gravitational settling is not included, the composition of the outer layers of the model remains constant and equal to the initial (i.e., early pre-main sequence) value until the onset of the first dredge-up.

By accounting for the differences in chemical composition between the stellar model and the RHD simulations, the correction factor $f_\mu$ in equation \eqref{mufactor} effectively replaces the entropy offset introduced in \citetalias{Spada_ea:2018} and \citetalias{Spada_Demarque:2019}.  
Although the mean molecular weight correction implemented in this work and the offset used in our earlier models are equivalent from a strictly operational standpoint, the current approach is to be preferred, because it is due to a well-understood physical mechanism, and it can be taken into account using a simple analytical expression.

\begin{figure}
\begin{center}
\includegraphics[width=0.49\textwidth]{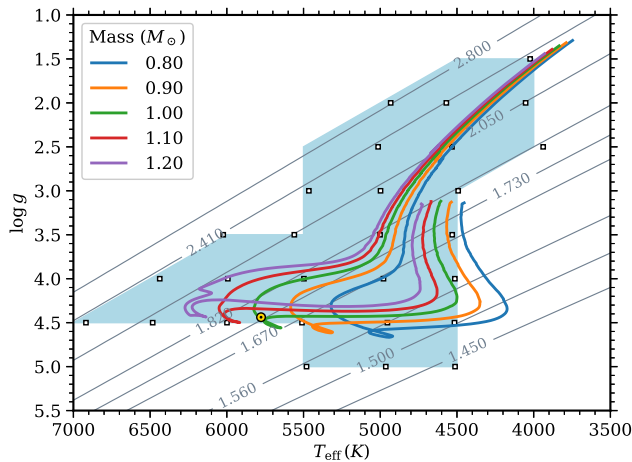}
\caption{Entropy-calibrated tracks in the Kiel diagram, calculated at solar metallicity and using the RHD simulations of \citet{Magic_ea:2013a}. The location of the RHD simulations (empty squares), and the domain of interpolation of the fitting function (light blue shading) are also shown. The grey contour lines represent the function $s_{\rm ad}^{\rm RHD}$ used in the calibration, in units of $10^9$ erg g$^{-1}$ $K^{-1}$. The location of the present Sun in the diagram is also marked by a yellow circle with a central dot.}
\label{kiel}
\end{center}
\end{figure}

\begin{figure*}
\begin{center}
\includegraphics[width=0.49\textwidth]{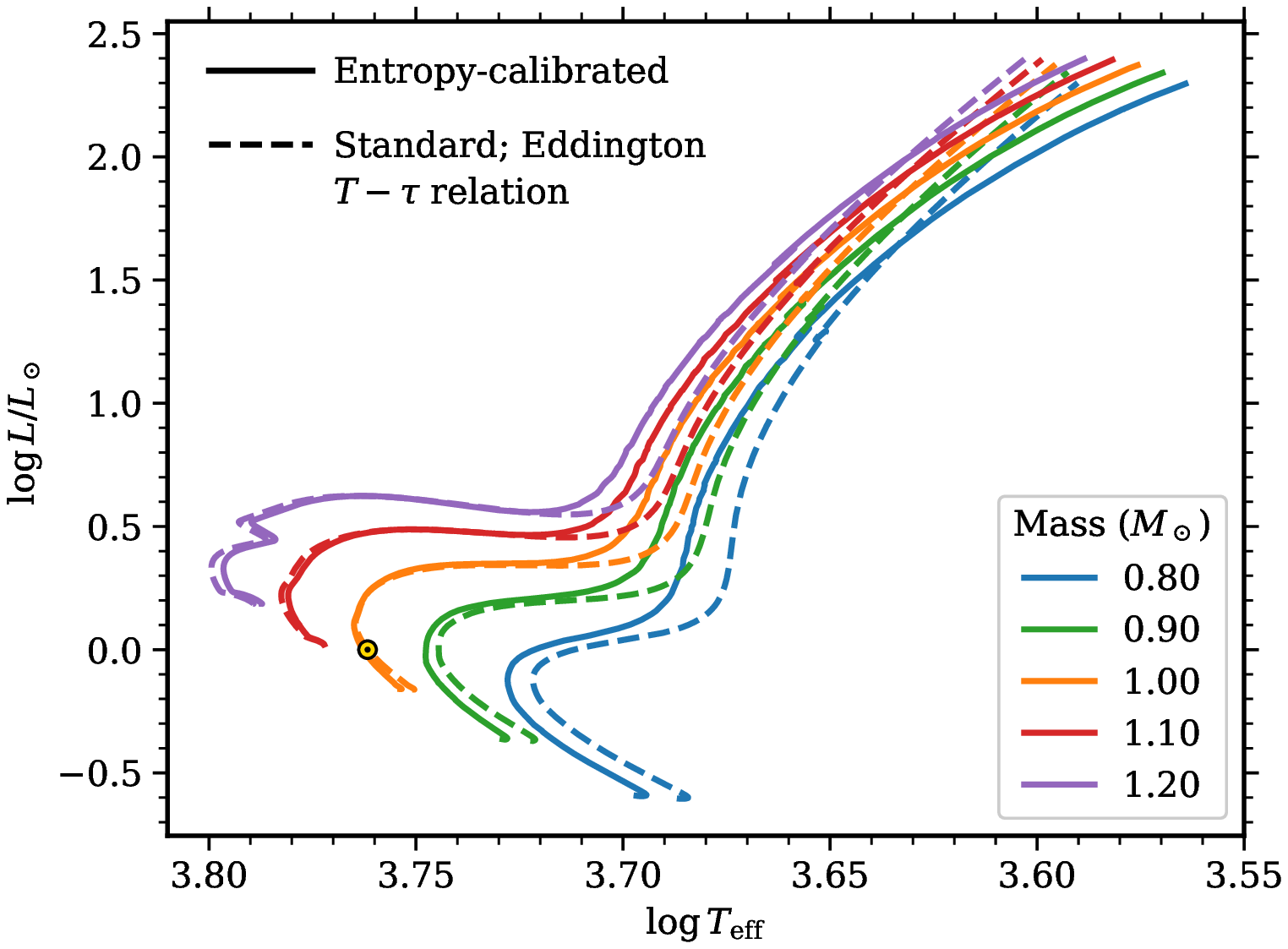}
\includegraphics[width=0.49\textwidth]{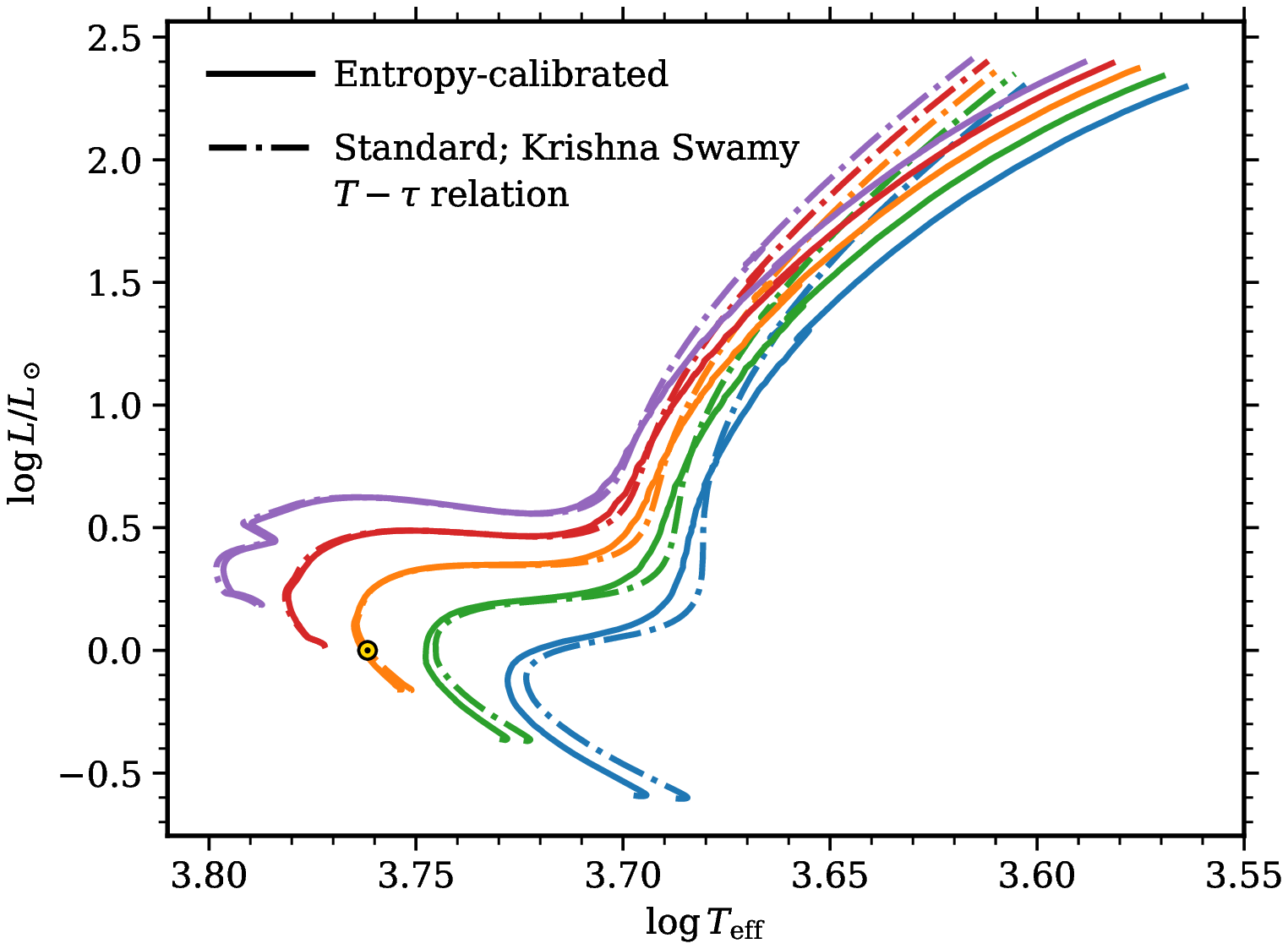}
\caption{Comparison of the entropy-calibrated tracks (solid lines) with standard ones (i.e., with constant, solar-calibrated \cmlt{}; dashed lines) in the HR diagram. Left panel: comparison with standard tracks implementing the Eddington grey $T$--$\tau$ relation, for which the solar-calibrated $\alpha_{\rm MLT} = 1.84$. Right panel: comparison with standard tracks implementing the Krishna Swamy $T$--$\tau$ relation, for which the solar-calibrated $\alpha_{\rm MLT} = 2.15$. In both panels, the position of the present Sun is marked by a yellow circle with a central dot.}
\label{tracks}
\end{center}
\end{figure*}

%%%%%%%%%%%%%%%%%%%%%%%%%%%%%%%%%%%%%%%%%%%%
\section{Results}
\label{results}
%%%%%%%%%%%%%%%%%%%%%%%%%%%%%%%%%%%%%%%%%%%%

\subsection{Entropy-calibrated evolutionary tracks in the Kiel and HR diagrams}

We constructed a set of entropy-calibrated evolutionary tracks extending from the pre-main sequence to the red giant branch past the so-called bump \citep{Thomas:1967}, up to $\log\, g \approx 1.5$.
The models have solar metallicity, initial helium fraction $Y=0.275$, and mass between $0.8$ and $1.2\, M_\odot$.
The entropy-calibrated tracks discussed in this Section were calculated using the fitting function of \citet{Magic_ea:2013a} to specify $s_{\rm ad}^{\rm RHD}$.
The models constructed with \sad{} specified from the \citet{Ludwig_ea:1999} fitting function are qualitatively similar, and the following discussion also applies to them.

The tracks are plotted in the ($T_{\rm eff}$, $\log\,g$), or Kiel diagram in Figure \ref{kiel}.
Note that the $1\, M_\odot$, entropy-calibrated track reproduces the present Sun.
In this sense, the correction factor $f_\mu$ implemented in our revised calibration (see Section \ref{factor}) is equivalent to the entropy offset correction used in \citetalias{Spada_ea:2018} and \citetalias{Spada_Demarque:2019}. 
Of course, to fully match the properties of the present Sun, the initial helium abundance also needs to be independently calibrated.
Except for part of the pre-main sequence, the tracks span the range of validity of the fitting function of \citet{Magic_ea:2013a}, effectively defined by the region of the diagram covered by their RHD simulations.
Outside this region, for instance for $\log\, g \lesssim 1.5$, although extrapolation of the fitting function is in principle possible, the fit is expected to quickly lose accuracy and meaning.

The two panels of Figure \ref{tracks} show a comparison in the HR diagram of the entropy-calibrated tracks with standard ones (i.e., using constant, solar-calibrated \cmlt{}), implementing the Eddington grey and the Krishna Swamy atmospheric $T$--$\tau$ relation, respectively.

It should be emphasised that, in contrast with the standard tracks, the entropy-calibrated tracks are utterly insensitive to the choice of the atmospheric $T$--$\tau$ relation, except, of course, for the actual numerical value of the calibrated \cmlt{} along the track.
This feature of the entropy calibration was first demonstrated in \citetalias{Spada_ea:2018} for the case of the solar model (see in particular Section 5 and Figure 11 of that paper).
We recovered this property in our current entropy-calibrated models, and we verified that it also applies to the red giant branch evolution.
More specifically, we have verified that entropy-calibrated tracks constructed with the Eddington grey or the Krishna Swamy $T$--$\tau$ relation, or even using surface boundary conditions given by the photospheric pressure derived from PHOENIX model atmospheres (see \citealt{Spada_ea:2017} for the details of the implementation in our stellar evolution code) are essentially indistinguishable from each other in all the evolutionary phases considered in this work (i.e., from early pre-main sequence to mature red giant branch).
For model details on this comparison, see Appendix \ref{atmosphere}.

For the main sequence and the sub-giant phase, the difference between the entropy-calibrated and standard tracks is larger, the lower the mass.
The location of the red giant branch, on the other hand, is shifted approximately equally for all the tracks.
With respect to the standard tracks implementing the Eddington grey $T$--$\tau$ relation, the shift is approximately $50$ K towards hotter effective temperatures between the base of the red giant branch and the bump (the tracks implementing the \sad{} from \citealt{Ludwig_ea:1999} feature a $T_{\rm eff}$ shift in the same direction, but larger, $\lesssim 100$ K). 
For the standard tracks implementing the Krishna Swamy $T$--$\tau$ relation, similar qualitative considerations apply.
Interestingly, these tracks are closer to the entropy-calibrated ones than the standard tracks implementing the Eddington $T$--$\tau$ relation.

Beyond the red giant branch bump, the entropy-calibrated tracks intersect and cross over the standard ones (both for the Eddington and the Krishna Swamy standard tracks).
It is not clear whether the different slope of the red giant branch in the HR diagram represents a real effect or it is an artefact due to the poor coverage of the upper region of the red giant branch by the RHD simulations.
Relatedly, a worse quality of the representation of the results of the RHD simulations by the fitting function at low $\log\, g$ is also a possibility.
It should be recalled that the fitting functions of \citet{Magic_ea:2013a} and \citet{Ludwig_ea:1999} also differ the most in this regime (see Figure \ref{sbot}).

Our results are in good qualitative agreement with those discussed in \citetalias{Spada_ea:2018} and \citetalias{Spada_Demarque:2019} in the pre-main sequence and main sequence portions of the tracks.
For the red giant branch, however, we find a large quantitative and qualitative disagreement: our previous results predicted a shift towards lower effective temperatures.
This discrepancy is a consequence of the inclusion of the correction factor discussed above (see Section \ref{factor}). 

Our current results are also in agreement with those of \citet{Salaris_Cassisi:2015}, and of \citet{Mosumgaard_ea:2020}.
Considering that the approaches used in those papers are largely independent from each other and from ours, this agreement is encouraging.

\begin{figure*}
\begin{center}
\includegraphics[width=0.49\textwidth]{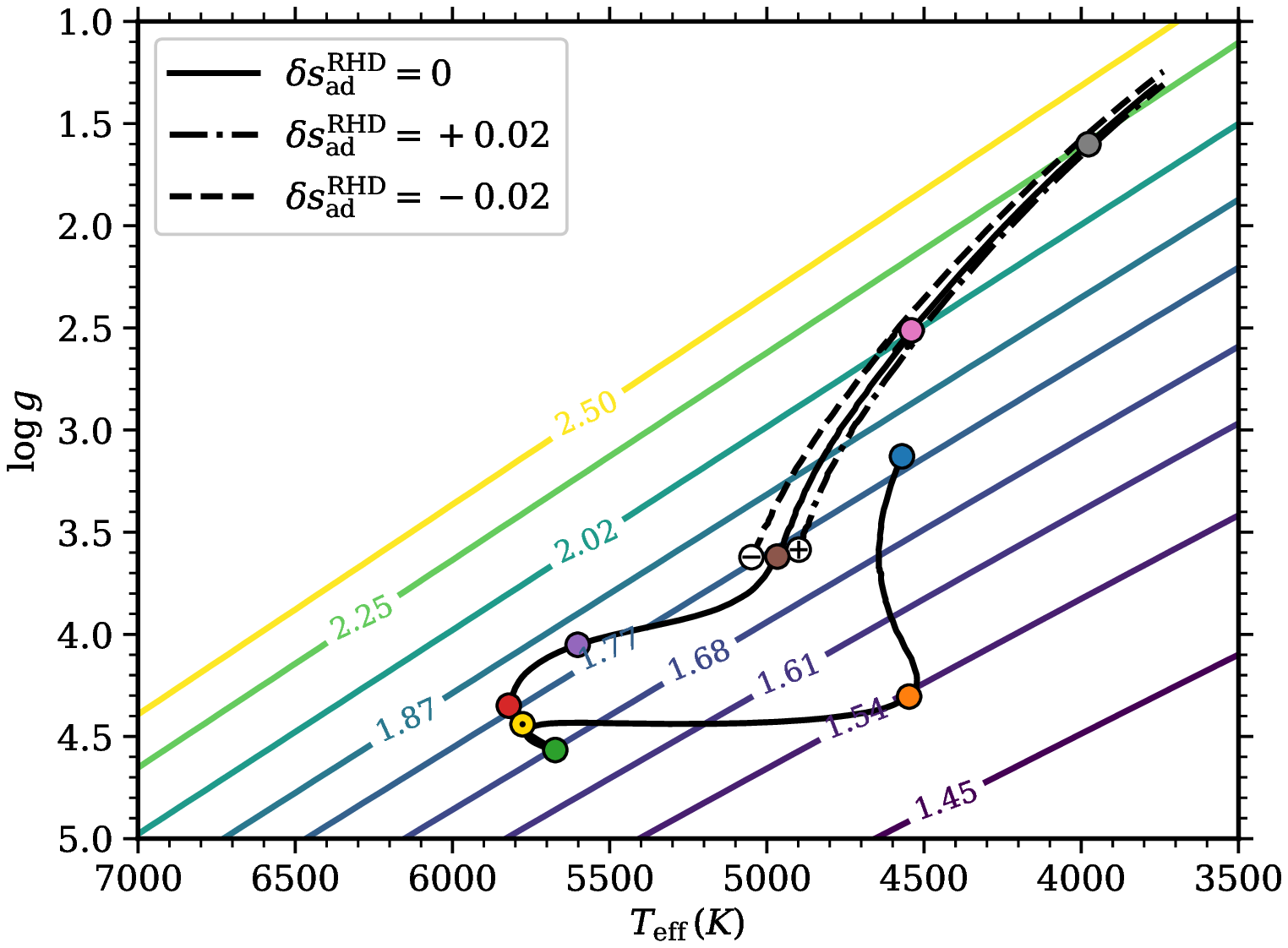}
\includegraphics[width=0.49\textwidth]{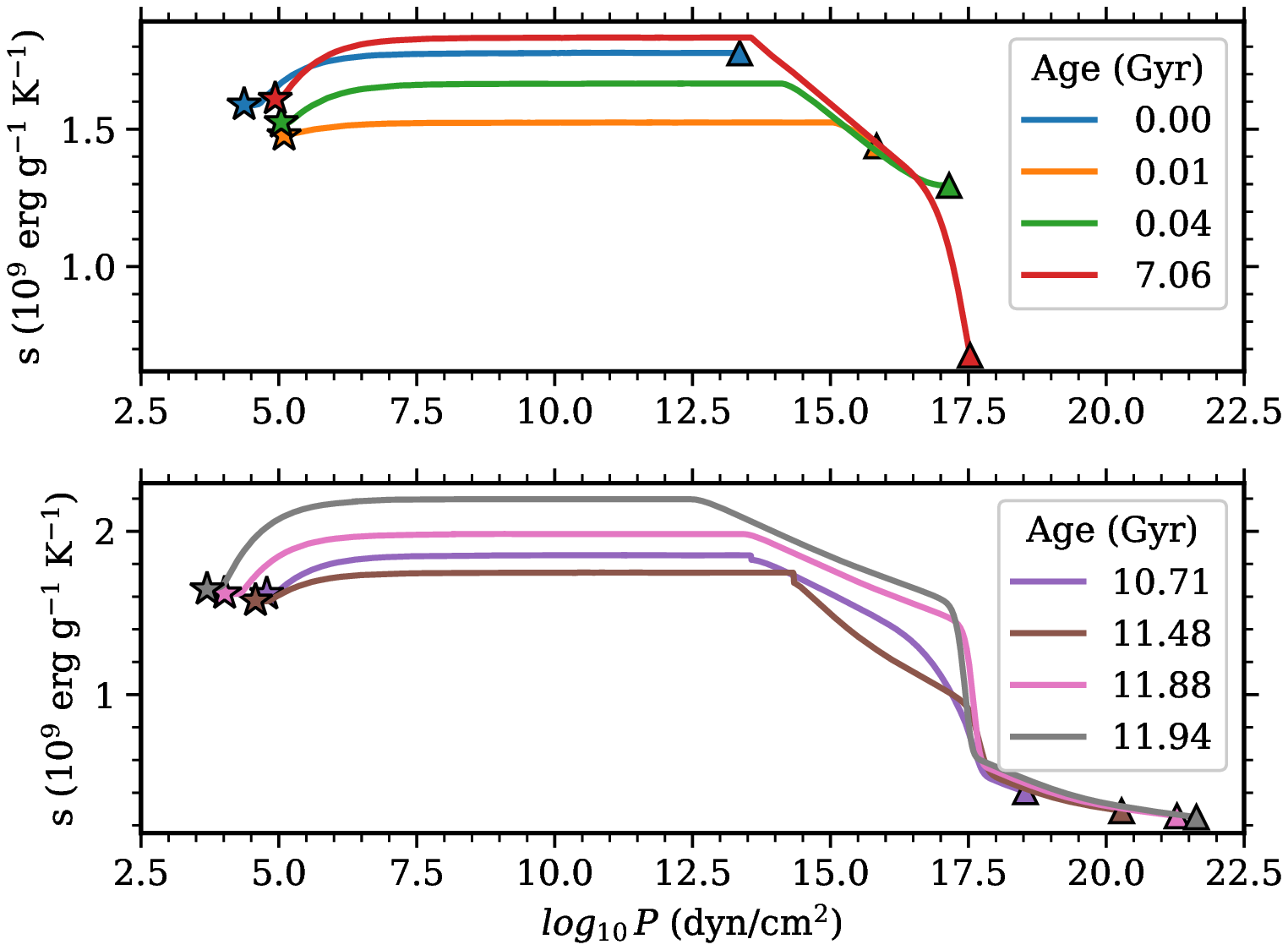}
\caption{Evolutionary track and interior entropy profiles at different ages for the $1\, M_\odot$ entropy-calibrated track. Left: Evolutionary track (solid black line), and level curves of adiabatic specific entropy from the fitting function of \citet{Magic_ea:2013a} used in the calibration; for illustrative purposes, two tracks calculated with an artificial offset added to the adiabatic specific entropy, $\delta s_{\rm ad}^{\rm RHD}$, are also shown (dot-dashed: positive offset; dashed: negative offset). All entropy values are quoted in units of $10^9$ erg g$^{-1}$ $K^{-1}$. The coloured circles mark the locations at which interior profiles are plotted in the panels on the right. The position of the present Sun is shown as a yellow circle with a central dot.
Right: interior profiles for the same track shown in the left panel; models up to the main sequence turn-off are shown in the top panel, while models on the sub-giant and red giant branch are plotted in the bottom panel. The triangles mark the centre of the model, while the stars mark the location of the photosphere.}
\label{entropy}
\end{center}
\end{figure*}

\begin{figure*}
\begin{center}
\includegraphics[width=0.49\textwidth]{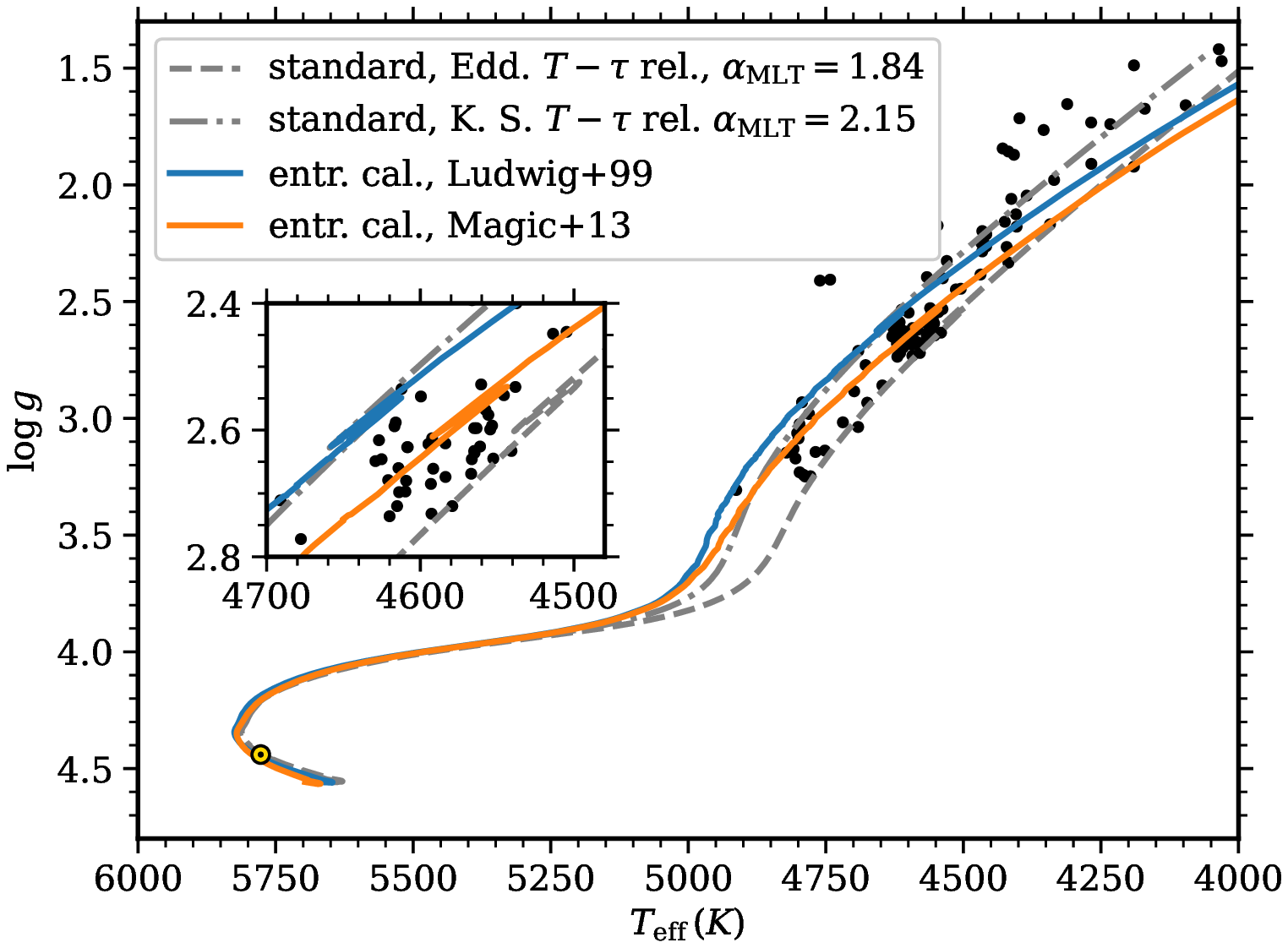}
\includegraphics[width=0.49\textwidth]{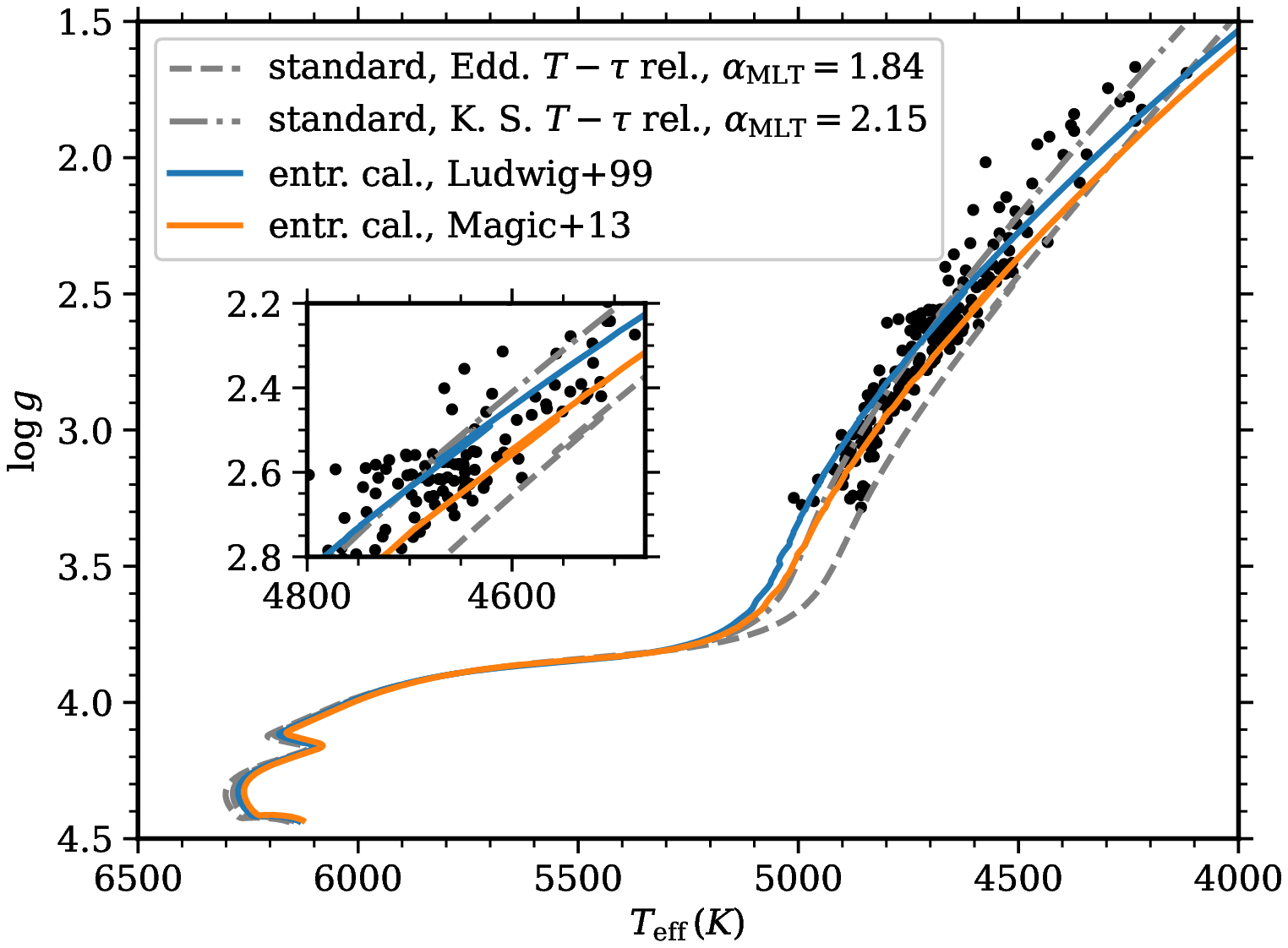}
\caption{Entropy-calibrated evolutionary tracks in the Kiel diagram compared with observational data for evolved stars.
Left panel: $1.0 \, M_\odot$; right panel: $1.2 \, M_\odot$. Data from the APOKASC2 catalog \citep{Pinsonneault_ea:2018}, selected at solar metallicity ($-0.1 < {\rm [Fe/H]} < +0.1$) and mass appropriate for each panel (within $\pm 0.05 \, M_\odot$ from that of the tracks). Two standard solar-calibrated tracks of the same mass, implementing the Eddington atmospheric $T$--$\tau$ relation (grey dashed line) and the Krishna Swamy one (grey dash-dotted line), are also shown for comparison. An enlargement of the area near the red giant bump is plotted in the insets in both panels. The position of the Sun is marked in the left panel by a yellow circle with a central dot.}
\label{apokasc2}
\end{center}
\end{figure*}

\subsection{Evolution of the interior entropy profile}
\label{interior_entropy}

To gain more insight into the physics of the entropy calibration, we analyse in detail the evolution of the entropy profile for the $1\, M_\odot$ track, illustrated in Figure \ref{entropy}.
Several interior profiles are shown in the two panels to the right, corresponding to the instants along the evolutionary track indicated by the coloured circles in the panel to the left.

The entropy profiles in the top-right panel correspond to pre-main sequence and main sequence models.
The specific entropy of the starting model is uniform except for the dip in the outermost layers, corresponding to the SAL, consistent with a fully convective structure.
In the next model, at the transition between the Hayashi and the Henyey track, an inner core in radiative equilibrium is present, where the entropy increases from the centre outwards.
The following profile, corresponding to the ZAMS model, is qualitatively similar, with a shallower convective envelope.
At the main sequence turn-off ($\approx 7$ Gyr), the entropy of the stellar core is significantly lower, due to the increase in mean molecular weight produced by the conversion of hydrogen into helium during the main sequence. 
We note that in the evolution from the birth-line to the end of the main sequence the entropy of the adiabatic region of the envelope first decreases, then increases, according to the prescription of the fitting function of the RHD simulations (contour lines in the left panel of Figure \ref{entropy}).

In the bottom-right panel, interior profiles corresponding to the post-main sequence are shown.
During the sub-giant phase, the adiabatic specific entropy is approximately constant, and then increases significantly during the red giant phase.
We also note that the entropy profiles feature a) a very steep profile in correspondence to the hydrogen burning shell ($\log_{\rm 10} P \approx 17.5$ in this model); b) a moderate, step-like discontinuity between the radiative interior and the convective envelope, due to the chemical composition gradient, which is smoothed out after the RGB bump (last two models); c) a steady increase of the entropy jump in the SAL, and decrease of the entropy at the centre, where the inert helium core is dominated by degeneracy. 
All these features of the entropy profile are qualitatively similar to those of a standard (constant \cmlt{}) model, the quantitative differences between the two being the result of the entropy calibration.

The left panel of Figure \ref{entropy} also shows two extra entropy-calibrated tracks, obtained by restarting the evolution from the base of the red giant branch and introducing an artificial specific entropy offset of $0.02\cdot 10^9$ erg g$^{-1}$ K$^{-1}$.
From their comparison with the unaltered track we see that a positive (negative) entropy offset produces a shift of the red giant branch towards lower (higher) effective temperature.
This behaviour can be understood in simple physical terms as follows.
At the leading order, an increase in the adiabatic specific entropy results in a model with a larger radius, as discussed, for instance, in Section 7.3.3 of \citet{HansenKawalerTrimble:2004}.
Since the entropy calibration leaves the luminosity of the star essentially unchanged \citepalias[see][]{Spada_ea:2018}, an increase in radius corresponds to a decrease in the effective temperature. 
This is a clear illustration of how the entropy budget of the star has a direct impact on its observable parameters.

\subsection{Radius and effective temperature calibration of red giants}

In practical terms, the entropy calibration yields a revised estimate of the radius and effective temperature of the stellar models. 
In agreement with our earlier results (\citetalias{Spada_ea:2018} and \citetalias{Spada_Demarque:2019}), we find that the luminosity is essentially unaffected.
The differences in radius and effective temperature with respect to standard models are modest during the late pre-main sequence, main sequence and sub-giant evolutionary phases, but they are quite substantial when the star is close to the Hayashi line, i.e., during the early pre-main sequence and the red giant branch phases (cf. Figure \ref{tracks}).

Our results are relevant, among other things, to the characterisation of red giant stars, and represent a testable prediction of the entropy calibration.
In Figure \ref{apokasc2} we compare our entropy-calibrated tracks with data from the APOKASC2 catalog \citep{Pinsonneault_ea:2018}. 
This catalog contains stellar properties for a large sample of $6676$ evolved stars, derived with a combination of ground based-spectroscopy (from the Apache Point Observatory Galactic Evolution Experiment, APOGEE), and Kepler asteroseismic data, analysed using five independent techniques.

Figure \ref{apokasc2} shows a subsample of stars from the APOKASC2 catalog selected at solar metallicity ($-0.1 < {\rm [Fe/H]} < +0.1$), and mass equal to $1.0 \pm 0.05 \, M_\odot$ (left panel) and to $1.2 \pm 0.05 \, M_\odot$ (right panel).
Together with the data, entropy-calibrated tracks of appropriate mass are plotted, implementing the $s_{\rm ad}^{\rm RHD}$ fitting function of both \citet{Ludwig_ea:1999} and \citet{Magic_ea:2013a}.
Standard tracks constructed with both the Eddington and the Krishna Swamy atmospheric $T$--$\tau$ relations are also plotted for for comparison.

With respect to the standard track with Eddington grey $T$--$\tau$ relation, the red giant branch of both the $1.0\, M_\odot$ and the $1.2\, M_\odot$ entropy-calibrated tracks is shifted to hotter effective temperatures by $\approx 50$ K ($\approx 100$ K for the \citealt{Ludwig_ea:1999} calibration), resulting in a better agreement with the data.
The standard track implementing the Krishna Swamy atmospheric $T$--$\tau$ relation, on the other hand, lies much closer to the entropy-calibrated ones.

It should be emphasised at this point that in the standard tracks, by construction, the parameter \cmlt{} has been calibrated to fit the present Sun. 
There is, however, no theoretical basis to justify that the solar-calibrated value of \cmlt{} also applies to the red giant branch evolution, and/or to a track of different mass.
The close agreement with the data of the standard tracks implementing the Krishna Swamy atmospheric $T$--$\tau$ relation thus comes at the price of one additional adjustable parameter with respect to the entropy-calibrated tracks, and could also be just a coincidence.

In fact, the solar calibration of the mixing length parameter is well-known to be sensitive, among other things, to the details of the treatment of the atmospheric layers, in particular to the $T$--$\tau$ relation, and to the optical depth at which the atmosphere is fitted to the model.
Such an interplay between \cmlt{} and the surface boundary conditions can result in large uncertainties in the effective temperature calibration of standard models \citep[see][]{Salaris_ea:2018, Choi_ea:2018}.
The sensitivity of our entropy-calibrated tracks to these modelling choices, in contrast, is minimal (see Appendix \ref{atmosphere} for details).

The Figure insets show an enlargement of the area around the red giant bump.
We note that while the effective temperature of the bump is shifted by the entropy calibration, its $\log\, g$ and luminosity are essentially unaffected.
Indeed, the luminosity and $\log \, g$ at which the bump occurs are essentially controlled by the evolution of the interior of the model (see, e.g., \citealt{Cassisi_ea:2002, Hekker_ea:2020} for detailed discussions), on which the entropy calibration has a modest impact. 

As discussed above (cf. Figure \ref{tracks}), the entropy-calibrated tracks intersect the standard one of the same mass above the red giant bump, at $\log \, g \approx 2$. 
This shallower slope of the red giant branch does not seem to be supported by the data, whose general trend is more compatible with that of the standard track.
We can interpret this as tentative evidence that the fitting functions (and/or the simulations from which they were derived) do not properly represent the evolution of the adiabatic specific entropy in the low effective temperature, low surface gravity regime. 
For instance, the assumption of plane-parallel geometry used in the simulations becomes increasingly questionable at low surface gravity.

In view of the discussion in Section \ref{interior_entropy}, the revision of the red giant branch location towards hotter effective temperatures implies that a lower value of $s_{\rm ad}$ in this phase is in better agreement with the observations.
In other words, a standard, solar-calibrated value of \cmlt{} produces red giant models with too high adiabatic specific entropy.

\subsection{Entropy-calibration of different convection formalisms: MLT vs CM}

Being part of the stellar evolution code, our entropy calibration procedure is not tied to the MLT, but can be seamlessly applied to a different convection theory.
In this Section we compare entropy-calibrated models implementing the MLT and the CM formalism.
The CM model was developed by \citet{Canuto_Mazzitelli:1991} to alleviate some of the shortcomings of the MLT, and it is the description of convection alternative to the MLT most commonly adopted in stellar evolution models.

Before delving into the comparison of the results, a brief discussion of the key differences between MLT and CM is in order.
In general, CM implements a sharper transition from efficient to inefficient convection. 
The physical necessity of this correction was argued for by \citet{Canuto:1996}.
As a result, CM predicts a much faster transition from radiative to adiabatic temperature stratification. 
This in turn leads to the radiative stratification holding to deeper layers when moving from the photosphere towards the interior, and thus to a higher superadiabaticity, followed by a quick transition to adiabatic stratification.
The overall effect is a narrower and higher SAL peak compared to that predicted by the MLT.
This behaviour is more or less emphasised in the different flavours of the CM theory (e.g., \citealt{Canuto_ea:1996}).

The CM entropy profile is closer than the MLT one to the results of RHD simulations; the converse is true, however, for the superadiabaticity profile (i.e., the $\nabla-\nabla_{\rm ad}$ profile; see, e.g., \citealt{Grimm-Strele_ea:2015}).
Assessing the relative merits of MLT and CM is a complex topic, that goes beyond the scope of the present work (see, e.g., the discussion in \citealt{Sonoi_ea:2019}). 
We note here that most of the success of the MLT is limited to the Sun, and coincidental\footnote{Assuming the same prescription for computing the mixing length scale, the steep superadiabatic gradients obtained with the CM approach are close to those found in RHD simulations for solar granules, while those from the MLT recover much better the average $\nabla-\nabla_{\rm ad}$ from the simulations. The fact, however, that neither CM nor MLT considers a separate modeling of up- and down-streaming flows makes any ``match'' of the RHD simulations coincidental.}, while the CM theory captures more closely the physics of turbulent convection. 
{
For instance, in the very shallow convection zones of A-stars, convective energy transport is everywhere inefficient, and the superadiabatic gradient is the same, and steep, in both up- and down-drafts (see, e.g., \citealt{Kupka_ea:2009}). 
As a consequence, using the CM approach is preferable in modelling these stars, while the MLT requires an ad-hoc, significantly non-solar value of \cmlt{} \citep{Smalley_Kupka:1997,Smalley_ea:2002}.
}

\begin{figure}
\begin{center}
\includegraphics[width=0.49\textwidth]{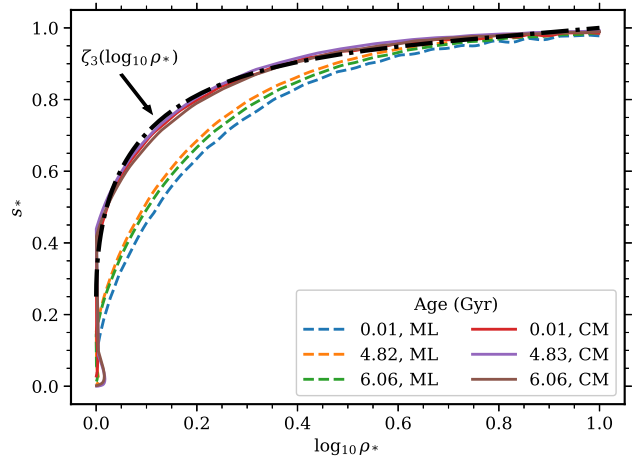}
\caption{Normalised entropy profile in the outer envelope of entropy-calibrated models implementing the CM and the MLT descriptions of convection (see equations \ref{smagic} for the definition of $s_*$ and $\rho_*$). The function $\zeta_3$ is the best fit derived by \citet{Magic:2016} to the normalised entropy profile from the RHD simulations of \citet{Magic_ea:2013a}.}
\label{interior}
\end{center}
\end{figure}

We compare entropy-calibrated models constructed with the CM vs the MLT description of convection in Figure \ref{interior}.
{We adopt the version of the CM model in which the scale length is given by the ``fully local" definition $\Lambda = \alpha_{\rm CM}\, H_P$, where $H_P$ is the local pressure scale height.}
In the Figure, scaled entropy profiles in the outer envelope at different ages are shown for a $1.2\, M_\odot$ model, chosen to be representative of the ZAMS, mid-main sequence, and red giant phases.
The scaling used in the plot is the same as in \citet{Magic:2016}. 
The normalised specific entropy $s_*$ and the normalised density $\rho_*$ are defined as:
\begin{equation}
\label{smagic}
s_* = \frac{s - s_{\rm min}}{s_{\rm ad}-s_{\rm min}}; \ \ \ \rho_* = \sqrt{\frac{\rho}{\rho_{\rm peak}}},
\end{equation}
where $s_{\rm min}$ is the minimum of the entropy, and $\rho_{\rm peak}$ is the density in correspondence of the SAL peak, i.e., where $\dfrac{ds}{dr}$ is maximum.  

The normalised entropy profiles are compared with the best-fit to the RHD simulations, given by the function $\zeta_3(\log_{\rm 10}\, \rho_*)$ defined in equation (7) of \citet{Magic:2016}.
The normalised entropy profiles of the models constructed with the CM theory of convection are remarkably close to the best-fit of the RHD simulations, represented by the function $\zeta_3$.
This is not the case for the models constructed using the MLT.
{This finding is consistent with previous work, and in particular with the comparison between the CM and the MLT presented in Figure 9 of \citet{Grimm-Strele_ea:2015} for the case of the Sun.}

Notably, the evolutionary tracks (not shown here) obtained with the MLT vs. CM descriptions of convection are essentially indistinguishable. 
This is consistent with the fact that the stellar parameters, i.e., radius and effective temperature, are mostly controlled by the value of the adiabatic specific entropy, and are much less sensitive to the details of the entropy profile.
This result lends support to the main theoretical foundation of the entropy calibration approach.

\subsection{Comparison with envelope-based and empirical \cmlt{} calibrations}

We emphasise once again that our approach focuses mainly on the adiabatic specific entropy and its relation to the radius and effective temperature of the model, while \cmlt{} is regarded as a purely numerical parameter of limited physical significance.
Nevertheless, comparison of our results with those of previous \cmlt{} calibrations from the literature may be of interest.

Among the previously published \cmlt{} calibrations, we make a distinction between ``envelope-based'' and ``empirical'' ones.
The former are constructed matching the value of \sad{} in a stellar envelope integration code with the results of RHD simulations \citep[e.g.][]{Ludwig_ea:1999,Magic_ea:2015a,Sonoi_ea:2019}.
The empirical calibrations, on the other hand, are obtained by running a stellar evolution code at constant \cmlt{} to match the observational parameters of a sample of stars \citep{Bonaca_ea:2012,Viani_ea:2018}.
In both cases, the calibration takes the form of polynomial functions of the independent variables $T_{\rm eff}$ and $\log\, g$ (and possibly metallicity).

\begin{figure}
\begin{center}
\includegraphics[width=0.49\textwidth]{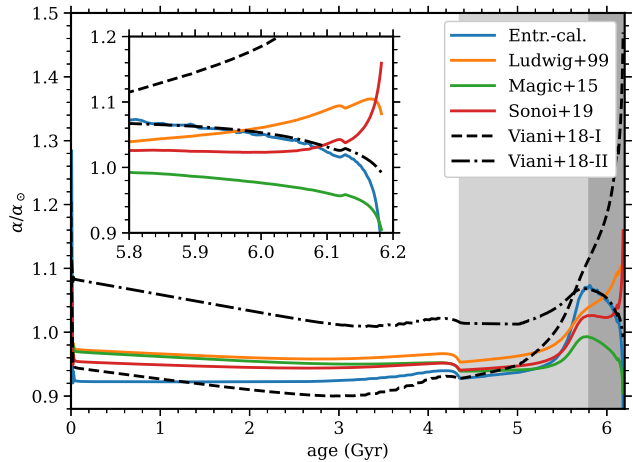}
\caption{Evolution of the mixing length parameter (scaled to the solar-calibrated value $\alpha_\odot$) according to our $1.2\, M_\odot$, solar metallicity tracks, compared with previous calibrations from the literature.
For the entropy-calibrated track, $\alpha_\odot = 1.88$ (see the text for details).
The inset focuses on the red giant branch phase (age $\gtrsim 5.8$ Gyr for this model).
The sub-giant and red giant phases are marked with a light and dark grey background, respectively.
Two empirical calibrations from \citet{Viani_ea:2018} are shown, obtained with different modelling choices: with core overshooting, but no gravitational settling (black dashed line, ``'Viani+18-I'); without core overshooting, but gravitational settling included (black dash-dotted line ``Viani+18-II'').}
\label{cmlt}
\end{center}
\end{figure}

Figure \ref{cmlt} shows the evolution of \cmlt{} for our $1.2\, M_\odot$ model calculated at solar metallicity; the Figure also shows \cmlt{} values from a selection of envelope-based and empirical calibrations applied to the same evolutionary track.
Given the well-known sensitivity of the MLT parameter to the details of the input physics and numerics of the stellar models, in order to obtain a meaningful comparison, we plot in each case the value of \cmlt{} scaled to the respective solar-calibrated value.
All the envelope-based and the empirical calibrations considered here provide directly this scaled value, $\alpha/\alpha_\odot$. 
For the entropy-calibrated models, we adopted $\alpha_\odot \approx 1.88$, obtained by interpolating \cmlt{} at $t_\odot = 4.57$ Gyr for our $1\, M_\odot$ entropy-calibrated track.
Two different empirical calibrations from \citet{Viani_ea:2018} are plotted in Figure \ref{cmlt}: their preferred one (hereafter ``Viani+18-I''), obtained with models implementing core overshooting, but no gravitational settling, and an alternative one (``Viani+18-II''), obtained with choices of input physics closer to ours (i.e., no core overshooting, but including gravitational settling).

Our results are in good qualitative and quantitative agreement with the envelope-based calibrations during the main sequence and sub-giant phases. 
The agreement deteriorates at the onset of the red giant branch phase (marked by the dark grey background in the main panel of the Figure), where the curves diverge significantly, as shown in more detail in the inset.
Although they differ quantitatively, our calibration and that of \citet{Magic_ea:2013a} feature a similar trend on the red giant branch phase.
This is not surprising, considering that they are derived from the same set of RHD simulations.

For the empirical calibrations, the evolution of $\alpha/\alpha_\odot$ predicted by the Viani+18-I fit is in reasonable agreement with that of our entropy-calibrated track during the main sequence and sub-giant phases; the agreement is completely lost on the red giant branch.
Interestingly, however, the Viani+18-II calibration, which is based on models with input physics closer to ours, is remarkably consistent with our entropy-based calibration on the red giant branch.

The results presented in this Section emphasise the strong dependence of \cmlt{} on the input physics in the models, even after a model-specific overall scaling (i.e., $\alpha_\odot$) has been accounted for.
Also, they show that the impact of different modelling choices on \cmlt{} may depend on the evolutionary phase under consideration (e.g., main sequence vs. red giant branch).

%%%%%%%%%%%%%%%%%%%%%%%%%%%%%%%%%%%%%%%%%%%%
\section{Discussion}
\label{discussion}
%%%%%%%%%%%%%%%%%%%%%%%%%%%%%%%%%%%%%%%%%%%%

In this paper we presented an implementation of the entropy calibration of the mixing-length parameter of convection that is fully integrated within the Yale stellar evolution code (see \citealt{Demarque_ea:2008}, \citealt{Spada_ea:2013,Spada_ea:2017} for details).
The method was first outlined theoretically by \citet{Tanner_ea:2016}.
\citetalias{Spada_ea:2018} and \citetalias{Spada_Demarque:2019} discussed a preliminary implementation, and demonstrated its application to modelling the Sun and $\alpha$ Centauri A and B, respectively.
The present paper, third in the series, contains significant methodological improvements, which greatly increase the overall performance, and extend the applicability of the entropy calibration to the red giant evolutionary phase.

This integrated entropy calibration of \cmlt{} is a promising avenue towards improving the description of convection in stellar evolution models on the basis of the more realistic results of the RHD simulations.
The main advantages of our approach are the following:
\begin{enumerate}
\item Ease of implementation: the entropy calibration is based on a single variable, \sad{}, whose functional dependence on the stellar parameters, equation \eqref{rhdentr}, can be expressed in analytical form;
\item Robustness: the relation between \sad{} and the surface parameters of the stars ($T_{\rm eff}$, $\log\,g$) is only moderately sensitive to the details of the atmosphere modeling, such as the $T$--$\tau$ relation, as was shown in \citetalias{Spada_ea:2018} and \citetalias{Spada_Demarque:2019};
\item Portability: since the calibration is built into the stellar evolution code, it allows effortless transition between different descriptions of convection, for instance, from MLT to CM.
\end{enumerate}

These advantages are to be weighted against the limitations of the methods.
Perhaps the most significant is renouncing to a detailed description of the stratification of the envelope provided by the RHD simulation, which is in contrast taken fully into account by the so-called on-the-fly patching method \citep{Mosumgaard_ea:2018,Mosumgaard_ea:2020}.
Secondly, the entropy calibration is of course only as accurate as the RHD simulations it is derived from \citepalias[cf. Section 3.2 of][]{Spada_Demarque:2019}.
These simulations are still computationally expensive, and extending and improving the existing grids is an onerous undertaking.
In particular, the relatively scarce coverage (and possibly lower accuracy) of the RHD simulations available to date for the upper red giant branch is the main limiting factor to the quality of the entropy-calibrated evolutionary tracks.

%%%%%%%%%%%%%%%%%%%%%%%%%%%%%%%%%%%%%%%%%%%%
\section{Conclusions}
\label{conclusions}
%%%%%%%%%%%%%%%%%%%%%%%%%%%%%%%%%%%%%%%%%%%%

We have described the implementation in a stellar evolution code of an improved calibration of the 1D convection theories used in stellar models, based on matching the adiabatic specific entropy derived from RHD simulations of convection. 
Our implementation is for the first time fully integrated within the flow of the stellar evolution code.
The enhanced internal consistency makes our code applicable to model stars beyond the main sequence.
Evolutionary tracks constructed with our method yield an improved consistency of the radii and effective temperatures of red giants stars with the available observational constraints.

\section*{Acknowledgements}
FS is supported by the German space agency (Deutsches Zentrum f\"ur Luft- und Raumfahrt) under PLATO Data Center grant 50OO1501.
FK is grateful to the Austrian Science Fund FWF for support through projects P29172-N and P33140-N and support from European Research Council (ERC) Synergy Grant WHOLESUN \#810218.
The authors are grateful to the referee for thoughtful comments which helped to improve the paper.

\section*{Data Availability}
The data underlying this article will be shared on reasonable request to the corresponding author.

\appendix

\section{Functional fits to the RHD simulations}
\label{rhdfits}

The functional fits used in this work to specify equation \eqref{rhdentr} have the following form.
\begin{itemize}
\item For the 2D RHD simulations of \citet{Ludwig_ea:1999}, valid for solar-metallicity: 
\begin{equation*}
f(\tilde T, \tilde g) = a_0 + a_5 \tilde T + a_6 \tilde g + a_1 \exp \left( a_3 \tilde T + a_4 \tilde g \right),
\end{equation*}
where $\tilde T=(T_{\rm eff}-5770)/1000$, $\tilde g = \log\, g - 4.4393$, and the coefficients $a_0 \dots a_6$ are given in Appendix B of \citet{Ludwig_ea:1999}.
\item For the 3D RHD simulations of \citet{Magic_ea:2013a}, valid for metallicity $-4.0 \leq {\rm [Fe/H]}\leq +0.5$:
\begin{equation*}
f(x,y,z) = \zeta_a(z) + \zeta_b(z) x + \zeta_c(z) y + \zeta_d(z)\, \exp \left[\zeta_e(z) x + \zeta_f(z) y \right],
\end{equation*}
where $x=(T_{\rm eff}-5777)/1000$, $y = \log\, g - 4.44$, $z={\rm [Fe/H]}$, and the functions $\zeta_i$ are second order polynomials in $z$, e.g., $\zeta_a(z) = a_0 + a_1 z + a_2 z^2$, and similar for $i=b,\dots,f$ (see Appendix B of \citealt{Magic_ea:2013a} for details). 
\end{itemize}

\section{Impact of the surface boundary conditions on the entropy-calibrated models}
\label{atmosphere}

In this appendix we explore in more detail the impact of the surface boundary conditions on the entropy-calibrated tracks.
We focus on two specific aspects, namely, the atmospheric $T$--$\tau$ relation, and the choice of $\tau_*$, the optical depth of the fitting point between the atmosphere and the model interior \citep[cf.][]{Salaris_ea:2018,Choi_ea:2018}.

\begin{figure}
\begin{center}
\includegraphics[width=0.5\textwidth]{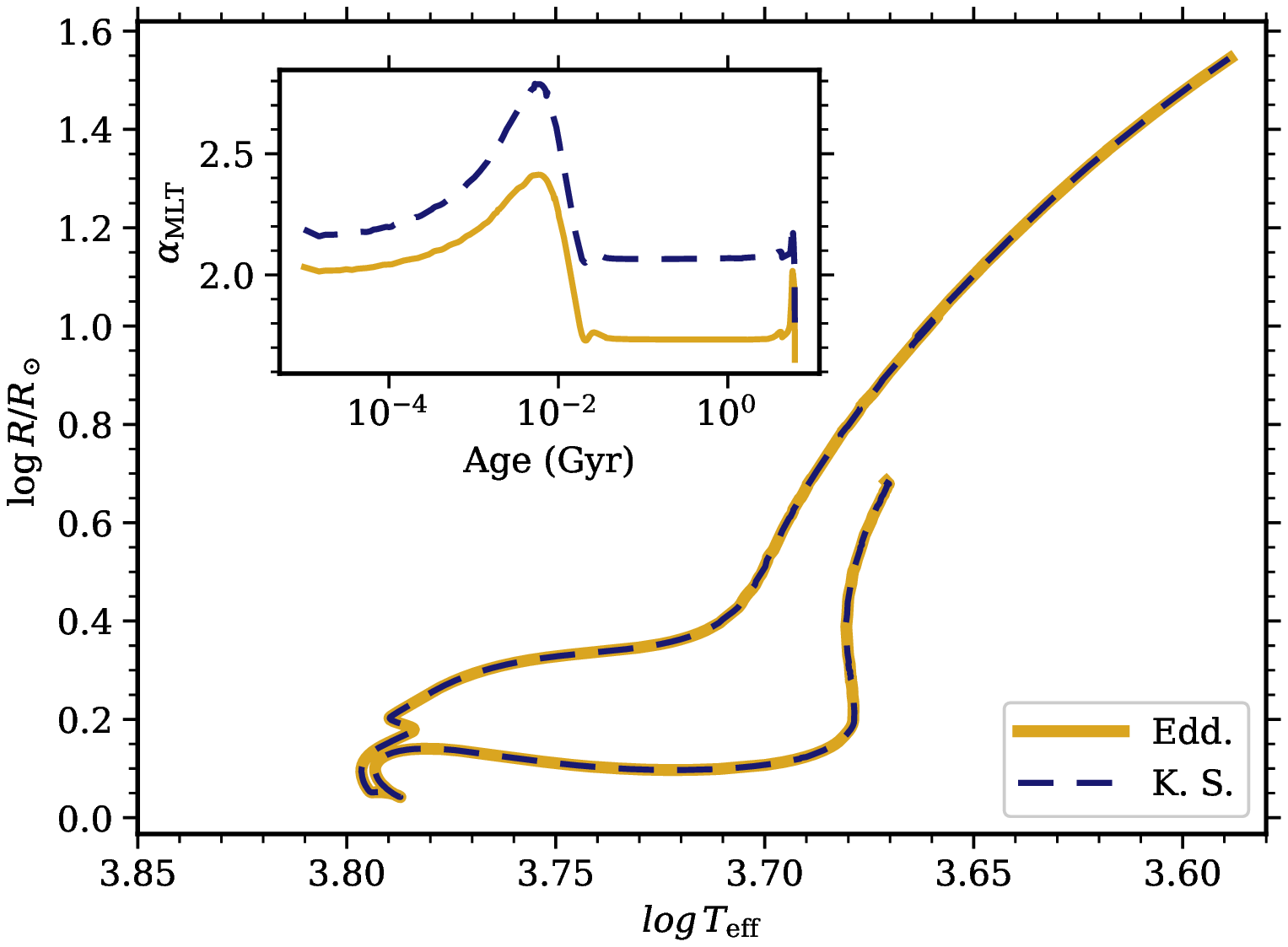}
\caption{Radius vs. effective temperature for entropy-calibrated tracks implementing different $T$--$\tau$ relations: Eddington grey (solid yellow), and Krishna Swamy (dashed blue). Both tracks have a mass of $1.2\, M_\odot$ and solar composition, and implement the RHD entropy fitting function of \citet{Magic_ea:2013a}. The inset shows the evolution of the entropy-calibrated \cmlt{} along each track.}
\label{T_tau}
\end{center}
\end{figure}

We compare entropy calibrated tracks constructed with different $T$--$\tau$ relations in Figure \ref{T_tau}.
The Figure shows two $1.2\, M_\odot$ tracks, calculated at solar metallicity, implementing the Eddington grey and the Krishna Swamy $T$--$\tau$ relations; the fitting function to the adiabatic specific entropy used in the calibration is that of \citet{Magic_ea:2013a}.
The atmospheric $T$--$\tau$ relation has a negligible impact on the radius and the effective temperature, and, in general, on the stellar parameters of our entropy-calibrated models. 
The only exception is the mixing length parameter itself, whose evolution is shown in the inset panel of the Figure.
This is of course a direct consequence of the entropy calibration procedure: a different atmospheric profile requires a different value of \cmlt{} to produce an envelope with the target adiabatic specific entropy, which is the same for both tracks.
The entropy-calibrated tracks are also essentially unchanged if the atmosphere integration is replaced by look-up tables providing the photospheric pressure derived from PHOENIX model atmospheres (not shown in the Figure). 
These results are fully consistent with \citetalias{Spada_ea:2018}, and show that the insensitivity to the $T$--$\tau$ relation also holds during the red giant branch evolution.

\begin{figure}
\begin{center}
\includegraphics[width=0.5\textwidth]{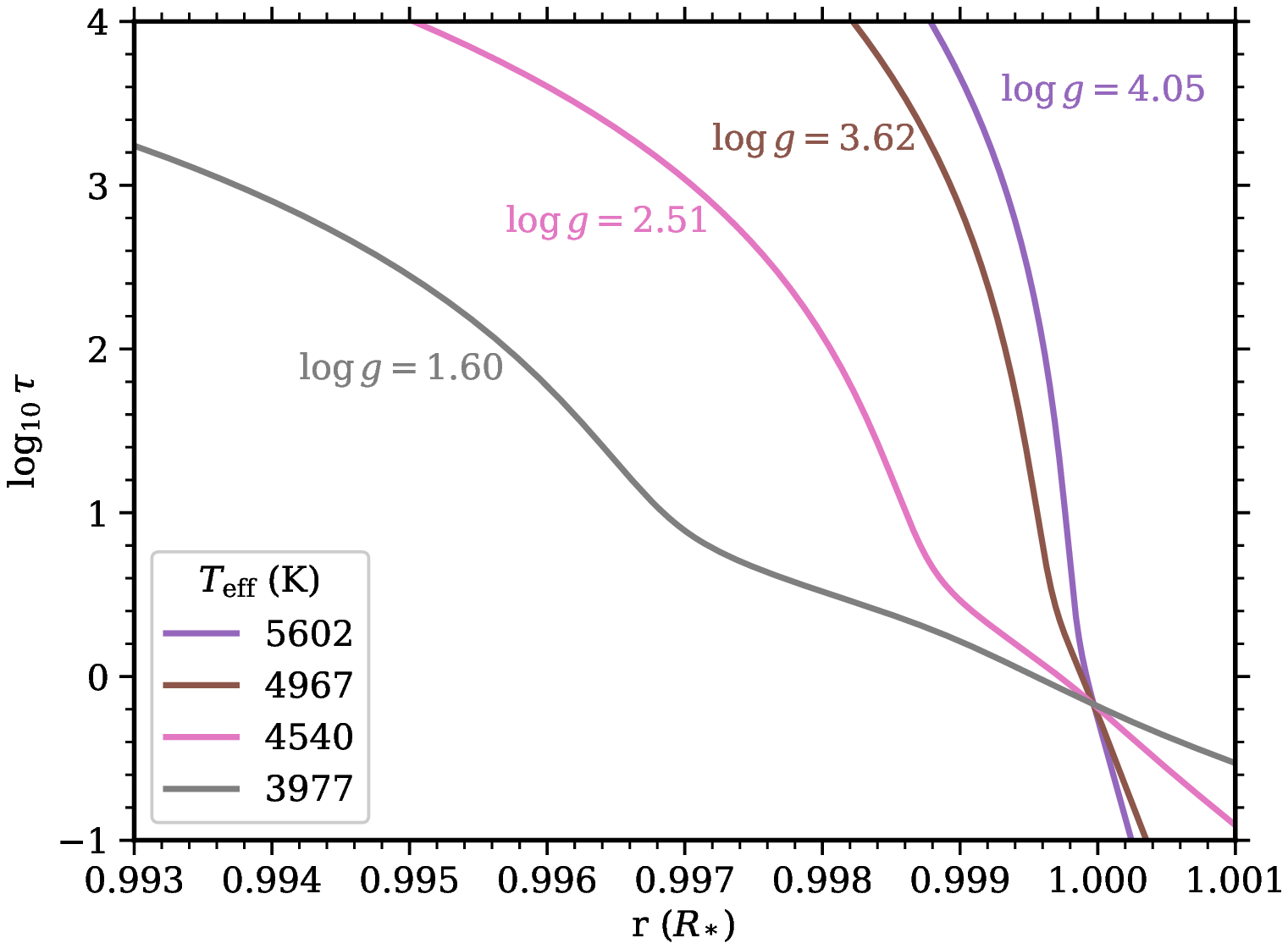}
\caption{Optical depth vs. fractional radius in the outer layers of entropy-calibrated red giant branch models ($1.0\, M_\odot$, solar composition, Eddington grey $T$--$\tau$ relation; cf. Figure \ref{entropy}). The photosphere is located at $\log_{10}\tau \approx - 0.1761$. The respective values of the surface gravity are also shown.}
\label{radius_tau}
\end{center}
\end{figure}

To quantify the uncertainty due to varying the optical depth $\tau_*$ at which the atmospheric fitting is performed, we proceed as follows.
Figure \ref{radius_tau} shows the optical depth vs. fractional radius in the outer layers for a selection of entropy-calibrated red giant branch models ($1.0\, M_\odot$, solar composition).
The models are the same as shown in the bottom right panel of Figure \ref{entropy}. 
Since the models implement the Eddington grey $T$--$\tau$ relation, their photosphere is located at $\log_{10} \, \tau = \log_{10} \frac{2}{3} \approx -0.1761$.
On the other hand, we can take $\tau = 10^2$ as a safe upper limit for a realistic choice of $\tau_*$.
We can therefore assume that the uncertainty in radius introduced by different choices of $\tau_*$ is bounded from above by:
\begin{equation*}
\frac{\Delta R}{R_*} = \left[ r(\log_{10 }\tau=-1) - r(\log_{10} \tau=3) \right] \lesssim 0.7\%,
\end{equation*}
where the numerical estimate is taken from Figure \ref{radius_tau}. 
We observe from the Figure that the uncertainty $\Delta R/R_*$ increases significantly as the star evolves up the red giant branch.
Assuming that the luminosity is unaffected by the changes in $\tau_*$, the radius uncertainty translates to an uncertainty in effective temperature of:
\begin{equation*}
\frac{\Delta T_{\rm eff}}{T_{\rm eff}} \approx \frac{1}{2} \frac{\Delta R}{R_*} \lesssim 0.35 \%.
\end{equation*}
We conclude that for the entropy-calibrated models considered here the uncertainty in effective temperature is $\approx 15$ K, or less, significantly less than for standard, solar-calibrated models.

\label{lastpage}
\end{document}